\documentclass[twocolumn,aps,prl,groupedaddress]{revtex4}
\usepackage{graphicx}
\usepackage{color}
\usepackage{amsmath}
\usepackage{enumitem}
\newcommand{\be}{\begin{equation}}
\newcommand{\ee}{\end{equation}}

\newcommand{\bea}{\begin{eqnarray}}
\newcommand{\eea}{\end{eqnarray}}

\newcommand{\p}{\partial}

\newcommand{\lb}{\left[}
\newcommand{\rb}{\right]}
\newcommand{\lp}{\left(}
\newcommand{\rp}{\right)}
\newcommand{\sgn}{{\rm sgn\,}}

\renewcommand{\Re}{{\rm \, Re\,}}

\renewcommand{\vec}[1]{{\bf #1}}

\begin{document}
\title{Electron viscosity, current vortices and negative nonlocal resistance in graphene}

\author{Leonid Levitov$^1$ and Gregory Falkovich$^{2}$}

\affiliation{$^1$ Department of Physics, Massachusetts Institute of Technology, Cambridge, Massachusetts 02139, USA}
\affiliation{$^2$ Weizmann Institute of Science, Rehovot 76100 Israel }
\affiliation{Institute for Information Transmission Problems, Moscow 127994 Russia}

% \date{\today}

\begin{abstract}
Quantum-critical strongly correlated electron systems are predicted to feature universal collision-dominated transport resembling that of viscous fluids.\cite{damle97,kovtun2005,son2007b,karsch2008}
However, investigation of these phenomena has been hampered by the lack of known macroscopic signatures of electron viscosity\cite{muller2009,mendoza2011,andreev2011,forcella2014,tomadin2014}.
Here we identify vorticity as such a signature and link it with a readily verifiable striking macroscopic DC transport behavior. Produced by the viscous flow, vorticity can drive electric current against an applied field, resulting in a negative nonlocal voltage. We argue that the latter may play the same role for the viscous regime as zero electrical resistance does for superconductivity. Besides offering a diagnostic which distinguishes viscous transport from ohmic currents, the sign-changing electrical response affords a robust tool for directly measuring the viscosity-to-resistivity ratio. Strongly interacting electron-hole plasma in high-mobility graphene \cite{gonzalez94,sheehy2007,fritz2008}
affords a unique link between quantum-critical electron transport and  the
%LL cornucopia
wealth of fluid mechanics phenomena. 
\end{abstract}

%\pacs{}
\maketitle
Symmetries and respective conservation laws play a central role in developing our understanding of strongly interacting states of matter.
%LL\cite{sachdev2001}.
%LL Symmetry approach proved particularly fruitful
This is the case, in particular, for many systems of current interest, ranging from quantum-critical states in solids and ultracold atomic gases to quark-gluon plasmas\cite{damle97,kovtun2005,son2007b,karsch2008}, which share common long-wavelength behavior originating from the fundamental symmetries of space-time.
The ensuing energy and momentum conservation laws
%LL Hydrodynamics
take the central stage in these developments,  defining hydrodynamics that reveals the universal collective behavior.
Powerful approaches based on conformal field theory and AdS/CFT duality
grant the well-established notions of fluid mechanics, such as  viscosity and vorticity,  an entirely new dimension\cite{sachdev2009,son2007}.

\begin{figure}
\includegraphics[scale=0.42]{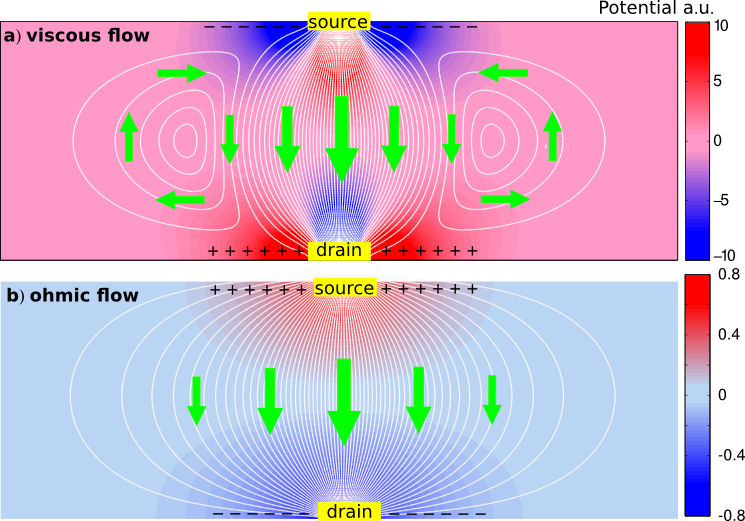}
\caption{{\bf Current streamlines and potential map for viscous and ohmic flows.} White lines show current streamlines, colors show electrical potential, arrows show the direction of current. Panel a) presents the mechanism of a negative electrical response: Viscous shear flow generates vorticity and a back flow on the side of the main current path, which leads to charge buildup of the sign opposing the flow and results in a negative nonlocal voltage. Streamlines and electrical  potential are obtained from Eq.\eqref{eq:psi(x,y)} and Eq.\eqref{eq:phi(x,y)}.
The resulting potential profile exhibits multiple sign changes and $\pm 45^{\rm o}$ nodal lines, see Eq.\eqref{eq:phi(x,y)_nodal_lines}. This provides directly measurable signatures of shear flows and vorticity.
Panel b) shows that,  in contrast, ohmic currents flow down the potential gradient, producing a nonlocal voltage in the flow direction.
}
\label{fig1}
%\vspace{-5mm}
\end{figure}

\begin{figure}
\includegraphics[scale=0.4]{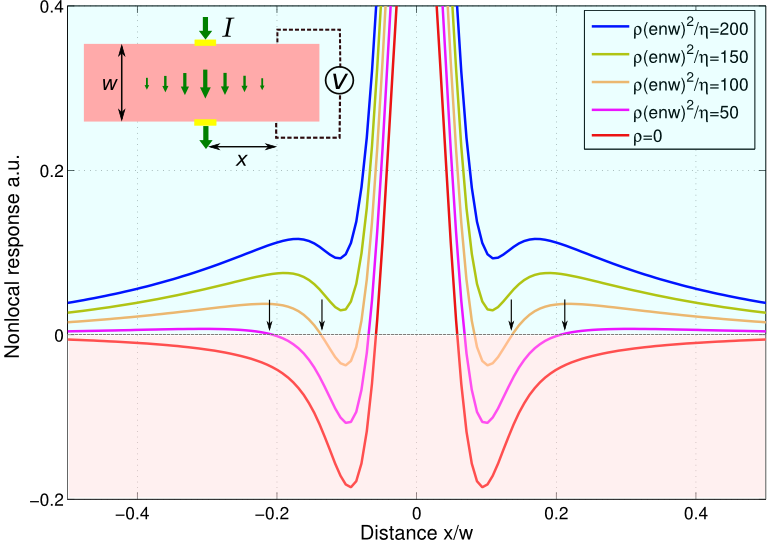}
\caption{{\bf Nonlocal response for different resistivity-to-viscosity ratios $\rho/\eta$.}
Plotted is voltage $V(x)$ at a distance $x$ from current leads obtained from Eq.\eqref{eq:V(x)_resistivity_nonzero} for the setup shown in the inset.
The voltage is positive in the ohmic-dominated region at large $|x|$ and negative in the viscosity-dominated region closer to the leads (positive values at even smaller $|x|$ reflect the finite contact size $a\approx 0.05 w$ used in simulation). Viscous flow dominates up to fairly large resistivity values, resulting in negative response
persisting up to values as large as $\rho(new)^2/\eta\approx 120$. Nodal points, marked by arrows, are sensitive to the $\rho/\eta$ value, which provides a way to directly measure viscosity (see text).
%\addLL{[Fix notation in legend]}
}
\label{fig2}
%\vspace{-5mm}
\end{figure}

Despite their prominence and new paradigmatic role, viscous flows in strongly correlated systems have so far lacked directly verifiable macroscopic transport signatures. Surprisingly, this has been the case even for
condensed matter systems
where a wide variety of experimental techniques is available to probe collective behaviors.
Identifying a signature that would do to viscous flows what zero electrical resistance did to superconductivity has remained an outstanding problem. The goal of this article is to point out that vorticity generated in viscous flows leads to a unique macroscopic transport behavior that can serve as an unambiguous diagnostic of the viscous regime. Namely, we predict that vorticity of the shear flows generated by viscosity can result in a backflow of electrical current that can {\it run against the applied field}, see Fig.\ref{fig1}. The resulting negative nonlocal voltage therefore provides a clear signature of the collective viscous behavior.
Associated with it are characteristic sign-changing spatial patterns of electric potential (see Fig.\ref{fig1} and Fig.\ref{fig2})
%LL \addLL{which are readily accessible with modern capacitance scanning microscopy techniques.\cite{yoo97,yacoby99}
which can be used to directly image vorticity and shear flows
%LL in electron systems
with modern scanning capacitance microscopy techniques.\cite{yoo97}
%LL,yacoby99}

The negative electrical response, which is illustrated in Fig.\ref{fig1}, originates from basic properties of shear flows.
We recall that the collective behavior of viscous systems results from
%LL particles rapidly exchanging
momenta rapidly exchanged in  carrier collisions while maintaining the net momentum conserved.
Since momentum remains a conserved quantity collectively, it gives rise to a hydrodynamic momentum transport mode.
%LL This mode describes
Namely, momentum flows in space, diffusing transversely to the source-drain current flow and away from the nominal current path. A shear flow established as a result of this process generates vorticity and (for an incompressible fluid) a back flow in the direction reverse to the applied field. Such a complex and manifestly non-potential flow pattern has a  direct impact on the electrical response, producing a reverse electric field acting opposite to the field driving the source drain-current (see Fig.\ref{fig2}).
This results in a negative nonlocal resistance which persists even in the presence of fairly significant ohmic currents (see Fig.\ref{fig2}).

Attempts to connect electron theory with fluid mechanics have
a long and interesting history,
partially summarized in Refs.\cite{dejong_molenkamp,jaggi91,forcella2014}. Early work on viscosity of Fermi liquids made connection with ultrasound damping.\cite{LifshitzPitaevsky_Kinetics} Subsequently, Gurzhi introduced an electronic analog of Poiseuille flow.\cite{gurzhi63} Related temperature dependent phenomena in nonlinear transport were observed by deJong and Molenkamp.\cite{dejong_molenkamp}
Recent developments started with the theory of a hydrodynamic, collision-dominated quantum-critical regime
%LL further developed
advanced by Damle and Sachdev.\cite{damle97}
%LL in the context of quantum-critical states of strongly-interacting systems.\cite{damle97}
%LL An important step was made by
% Andreev {\it et al.}
Andreev, Kivelson, and Spivak
%LL \cite{andreev2011} who
argued that hydrodynamic contributions can dominate resistivity in systems with a large disorder correlation length.\cite{andreev2011}
Forcella {\it et al.} predicted that electron viscosity can impact electromagnetic field penetration in a dramatic way.\cite{forcella2014} Davison {\it et al.} linked electron viscosity to
linear resistivity of the normal state of the copper oxides.\cite{davidson2014}

As a parallel development, recently there was a surge of interest in electron viscosity of graphene.\cite{muller2009,mendoza2011,tomadin2014,principi2015,cortijo2015}
The quantum-critical behavior is predicted to be particularly prominent in graphene.\cite{gonzalez94,sheehy2007,fritz2008}
Electron interactions in graphene are strengthened near charge neutrality (CN) due to the lack of screening at low carrier densities.\cite{kashuba08,fritz2008} As a result, carrier collisions are expected to dominate transport  in pristine graphene in a wide range of temperatures and dopings.\cite{narozhny2015} Furthermore, estimates of electronic viscosity near CN yield one of the lowest known viscosity-to-entropy ratios which approaches the universal AdS/CFT bound.\cite{muller2009}

Despite the general agreement that graphene holds the key to electron viscosity, experimental progress has been hampered by the lack of easily discernible signatures in macroscopic transport. Several striking effects have been predicted, such as vortex shedding
%and associated narrow-band noise
in the preturbulent regime induced by strong current\cite{mendoza2011}, as well as nonstationary
% the interplay of the azimuthal and radial
flow in a `viscometer' comprised of an AC-driven Corbino disc.\cite{tomadin2014}
% was proposed
These proposals, however, rely on fairly complex AC phenomena originating from high-frequency dynamics in the electron system.
In each of these cases, as well as in those of Refs.\cite{forcella2014,davidson2014},
%LL however,
a model-dependent analysis was required to delineate the effects of viscosity from `extraneous' contributions. In contrast, the nonlocal DC response considered here is a direct manifestation of the collective momentum transport mode which underpins viscous flow, therefore providing an unambiguous, almost textbook, diagnostic of the viscous regime.

Nonlocal electrical response mediated by chargeless modes was found recently to be uniquely sensitive to the quantities which are not directly accessible in electrical transport measurements, in particular spin currents and valley currents.\cite{abanin2011a,abanin2011b,gorbachev2014}
In a similar manner, the nonlocal response discussed here gives a diagnostic of viscous transport, which is more direct and powerful than any approaches based on local transport.

There are several aspects of the electron system in graphene that are particularly well suited for studying electronic viscosity. First, the momentum-nonconserving Umklapp processes are forbidden in two-body collisions because of graphene crystal structure and symmetry.
% This ensures momentum conservation and underpins the prominence of the
This ensures the prominence of momentum conservation and associated
collective
%hydrodynamical momentum
transport. Second, while carrier scattering is weak away from charge neutrality, it can be enhanced by several orders of magnitude by tuning the carrier density to the neutrality point.
%Second, while electron collision rate is small away from charge neutrality,
%% (or, Dirac point),
%by tuning the carrier density to the neutrality point the collision rate can be enhanced by several orders of magnitude. 
This allows to cover the regimes of high and low viscosity, respectively, in a single sample.  Lastly, the two-dimensional structure and atomic thickness makes electronic states in graphene fully exposed and amenable to sensitive electric probes.

To show that the timescales are favorable for the hydrodynamical regime, we will use parameter values estimated for pristine graphene samples which are almost defect free, such as free-standing graphene.\cite{bolotin2008}
Kinematic viscosity can be estimated as the momentum diffusion coefficient
$\nu\approx\frac12 v_F^2\gamma_{\rm ee}^{-1}$ where $\gamma_{\rm ee}$ is the carrier-carrier scattering rate, and $v_F=10^6\,{\rm m/s}$ for graphene. According to Fermi-liquid theory, this rate behaves as $\gamma_{\rm ee}\sim (k_{\rm B}T)^2/E_F$ in the degenerate limit i.e. away from charge neutrality, which leads to large $\nu$ values. Near charge neutrality, however, the rate $\gamma_{\rm ee}$ grows and $\nu$ approaches the AdS/CFT limit, namely $s \hbar /4\pi k_{\rm B}$ where $s$ is entropy density. Refs.\cite{kashuba08,fritz2008} estimate
this rate as $\gamma_{\rm ee}\approx A \alpha^2 k_{\rm B}T/\hbar$, where  $\alpha$ is the interaction strength. For $T=100\,{\rm K}$, assuming $E_F=0$ and approximating the prefactor as $A\approx 1$\cite{kashuba08,fritz2008}, this predicts  characteristic times as short as $\gamma_{\rm ee}^{-1} \approx 80 \, {\rm fs}$.
Disorder scattering can be estimated from the measured mean free path values which reach a few microns at large doping \cite{thiti13}. Using the momentum relaxation rate square-root dependence on doping, $\gamma_p
\propto n^{-1/2}$, and
%LL extrapolating to
estimating it near charge neutrality, $n\lesssim 10^{10}\,{\rm cm^{-2}}$, gives times $\gamma_p^{-1}\sim 0.5\,{\rm ps}$, which are longer than the values $\gamma_{\rm ee}^{-1}$ estimated above.
The inequality $\gamma_p\ll \gamma_{\rm ee}$ justifies our hydrodynamical description of transport.

Momentum transport in the hydrodynamic regime is described by continuity equation for momentum density,
\be
\p_t p_i+\p_j T_{ij}=-\gamma_p p_i
,\quad
T_{ij}=P\delta_{ij}+\mu v_iv_j+T_{ij}^{\rm (v)}
,
\ee
where $T_{ij}$ is the
%LL so-called stress tensor describing
momentum flux tensor, $P$ and $\mu$ are pressure and mass density,  and $\vec v$ is the carrier drift velocity. The quantity $\gamma_p$, introduced above, describes electron-lattice momentum relaxation due to disorder or phonons, which we will assume to be small compared to the carrier scattering rate.
We can relate pressure to the electrochemical potential via $P=e\int_{n_0}^n \Phi(n')dn'$. Here we work at degeneracy, $E_F\gg k_{\rm B} T$, ignoring the entropic/thermal contributions, and approximating $P\approx e(n-n_0)\Phi$, with $n$ the particle number density. While carrier scattering is suppressed at degeneracy as compared to its value at $E_F=0$, here we assume that the carrier-carrier scattering remains faster than the disorder scattering, as required for the validity of hydrodynamics. Viscosity contributes to the momentum flux tensor through
%LL For systems at degeneracy, we can relate pressure to the electrochemical potential via $P=en\Phi$ with $n$ the particle number density. Viscosity contributes to the stress tensor via
\be
T_{ij}^{\rm (v)}=\eta(\p_iv_j+\p_jv_i)+(\zeta-\eta)\p_kv_k\delta_{ij}
\ee
where $\eta$ and $\zeta$ are the first and second viscosity coefficients.
%LL When  drift velocities are
For drift velocities smaller than plasmonic velocities, transport in charged systems is described by an incompressible flow with a divergenceless velocity field,
%LL, whose velocity field is divergenceless,
$\p_i v_i=0$. In this work, we consider the limit of low Reynolds number,
%LL when
$\mu v_iv_j\ll\eta(\p_iv_j+\p_jv_i)$,
%. In this limit,
such that the role of viscosity is most prominent.
%LL ; for potential ways of detecting signatures of hydrodynamic behavior at finite Reynolds numbers, see \cite{mendoza2011}.
At linear order in $\vec v$, we obtain an electronic Navier-Stokes equation
%LL (NS)  equation
\be\label{eq:NS}
\p_tp_i-\eta\nabla^2 v_i +\gamma_p p_i=-\p_i P
.
\ee
This equation describes momentum transport: imparted by the external field $\vec f=-\nabla P$, momentum flows to system boundary where dissipation takes place.
%($\gamma_p$ accounts for residual disorder scattering in the bulk).
It is therefore important to endow Eq.\eqref{eq:NS} with suitable boundary conditions. In fluid mechanics this is described by the no-slip boundary condition $\vec v=0$. We use a slightly more general boundary condition
%LL It is instructive to generalize this condition to account for partial slippage:
\be\label{eq:b.c.}
v_\perp =0,\quad
v_\parallel=-\alpha \p_\parallel P
\ee
where the subscripts $\perp$ and $\parallel$ indicate the velocity and derivative components normal and tangential to the boundary. The second relation in Eq.\eqref{eq:b.c.} generalizes the no-slip condition to account for non-hydrodynamical effects in the boundary layer on the scales $\gtrsim l=v/\gamma_{\rm ee}$. The model in Eq.\eqref{eq:b.c.}, equipped with the parameter $\alpha$,  provides a convenient way to assess the robustness of our predictions.
%LL The width of the boundary layer is on the order of the carrier mean free path $l=v/\gamma_{\rm ee}$.

%LL As a quick illustration,
It is instructive to consider current flowing down a long strip of a finite width. A steady viscous flow features a nonuniform profile in the strip cross-section governed by the momentum flow to the boundary.  Eq.\eqref{eq:NS}, applied to a strip $0<y<w$, yields $(-\eta \p_y^2+\gamma_p m n)v(y)=enE$, where $v(y)$ and $E$ are the drift velocity and electric field directed along the strip (and $m$ is an effective mass defined through the relation $\vec p=m n \vec v$). Setting $\alpha$ and $\gamma_p$ to zero for simplicity, we find a parabolic profile $v(y)=Ay(w-y)$, where $A=neE/2\eta$ and $\eta=mn\nu$. The nonzero shear $\p_yv=A(w-2y)$ describes momentum flow to the boundary. The net current $I=\int_0^w nev(y')dy'=(n^2e^2/12\eta) w^3 E$ scaling as a cube of the strip width is the electronic analog of the Poiseuille law. Being distinct from the linear scaling $I\propto wE$ in the ohmic regime, the cubic scaling can in principle be used to identify the viscous regime. It is interesting to put the current-field relation in a ``Drude'' form using kinematic viscosity: $I=\frac{ne^2\tau_w}{m}wE$ with $\tau_w=w^2/12\nu$ an effective scattering time. Evaluating the latter as $\tau_w\approx\frac16(w/v_F)^2\gamma_{\rm ee}$ we find values that, for realistic system parameters,  can greatly exceed the naive estimate $\gamma_p^{-1}=w/v_F$ based on the ballistic transport picture. This remarkable observation was first made by Gurzhi~\cite{gurzhi63}.

Next, we proceed to analyze nonlocal response in a strip with transverse current injected and drained through a pair of contacts as pictured in Fig.\ref{fig1}.
Unlike the above case of longitudinal current, here the potential profile is not set externally but must be obtained from \eqref{eq:NS}. The analysis is facilitated by introducing a stream function through 
$\vec v=\vec z\times\nabla\psi$, which solves the incompressibility condition. At first we will completely ignore the ohmic effects, setting $\alpha$ and $\gamma_p$ to zero as above, which leads to a biharmonic equation $\lp\p_x^2+\p_y^2\rp^2\psi=0$ with the boundary conditions $v_x=0$, $nev_y=I\delta(x)$ for $y=0,w$.  Using Fourier transform in $x$, we write $\psi(x,y)=(2\pi)^{-1}\int dk e^{ikx}\psi_k(y)$ and then determine $\psi_k(y)$ separately for each $k$ (see Supplementary Information). Inverting Fourier transform gives the stream function
\bea\label{eq:psi(x,y)}
\psi(x,y)=&&  \frac{I}{ne}  \int \frac{dke^{ikx}}{2\pi i k}\left( \frac{e^{ky}+e^{k(w-y)}}{e^{kw}+1}\right.
\\ \nonumber
&&
\left. +a_k [y\sinh k(w-y) + (w-y)\sinh ky]\right)
,
\eea
where we defined $a_k=k\tanh (kw/2)/(kw + \sinh kw)$.  Contours (isolines) of $\psi$ give the streamlines for the flow shown in Fig.\ref{fig1}. While most of them are open lines connecting source and drain, some streamlines form loops. The latter define vortices occurring on both  sides of the current path. Numerically we find that vortex centers are positioned very close to $x=\pm w$ (see Supplementary Information).

We can now explore the electrical potential of the viscous flow. The latter can be found
directly from $\psi(x,y)$ giving
\be\label{eq:phi(x,y)}
\phi(x,y)= \frac{\beta I}2 \int dke^{ikx} a_k
[\sinh k(y-w)+\sinh ky]
,
\ee
where we defined $\beta=2\eta /(\pi n^2 e^2)$ (see Supplementary Information). As illustrated in Fig.\ref{fig1}, Eq.\eqref{eq:phi(x,y)} predicts a peculiar sign-changing spatial dependence, with two pairs of nodal lines crossing at contacts.
 To understand this behavior, we evaluate $\phi(x,y)$ explicitly
in the regions near contacts $(x,y)=(0,0),(0,w)$. Near the first contact, approximating $\tanh (kw/2)\approx \sgn k$, $\sinh ky\approx \frac12 e^{|k|y}\sgn k$, etc, we find
\be\label{eq:phi(x,y)_nodal_lines}
\phi(x,y)\approx
\frac{\beta I}2\int dke^{ikx} |k| e^{-|k|y}
= \frac{\beta I (y^2-x^2)}{(y^2+x^2)^2}
\ee
($|x|,|y|\ll w$). Eq.\eqref{eq:phi(x,y)_nodal_lines} predicts an inverse-square dependence vs. distance from contacts and also the presence of two nodal lines running at $\pm 45^{\rm o}$ angles relative to the nominal current path. Similar behavior is found near the other contact, $\phi(x,y)\approx -\frac{\beta((w-y)^2-x^2)}{((w-y)^2+x^2)^2}I$. We note that the $r^{-2}$ power law  dependence is much stronger than the $\ln r$ dependence expected in the ohmic regime. This, as well as multiple sign changes, provides a clear signature of a viscous flow.

The nonlocal voltage measured at a finite distance from the current leads (see schematic in Fig.\ref{fig2} inset) can be evaluated as $V(x)=\phi(x,w)-\phi(x,0)$. From Eq.\eqref{eq:phi(x,y)_nodal_lines} we predict
voltage that is falling off as $x^{-2}$ and is of a {\it negative sign}:
\be\label{eq:R(x)=-1/x^2}
V(x)\approx -\frac{2\beta}{x^2}I
\ee
($|x|\lesssim w$). Microscopically, negative voltage originates from a viscous shear flow which creates vorticity  and  backflow on both sides of the current path, see Fig.\ref{fig1}.

Numerically we see that the negative response persists to arbitrarily large distances, see $\rho=0$ curve in Fig.\ref{fig2}. The sign change at very short $x$, evident in Fig.\ref{fig2}, arises due to a finite contact size. We model it by replacing the delta function in the boundary condition for current source by a Lorentzian, $nev_y=I {a}/{\pi (x^2+a^2)}$ at $y=0,w$. After making appropriate changes in the above derivation (namely, plugging $e^{-a|k|}$ under the integral) we find
\be\label{eq:R(x)_sign_change}
V(x)\approx -\frac{\beta I}{(x-ia)^2}+{\rm c.c.}
=-\frac{2\beta I (x^2-a^2)}{ (x^2+a^2)^2}
\ee
This expression exhibits a sign change at $x=a$ (representing ``the contact edge'') and is negative for all $|x|>a$, i.e. everywhere outside contacts (this is further discussed in Supplementary Information).

It is interesting to probe to what extent the negative response is sensitive to boundary conditions, in particular to the no-slip assumption. Extending the above analysis to the
%LL partial-slip
boundary conditions with nonzero $\alpha$ in Eq.\eqref{eq:b.c.} we find the nonlocal response of the form
\be
V(x) =
\beta I \int dke^{ikx} \frac{k\tanh (kw/2)\sinh kw}{ kw+(1+\tilde\alpha k^2)\sinh kw}
,
\ee
where $\tilde\alpha=2\eta\alpha/ne$.
%\addLL{[check factors, write dimensional estimate for $\alpha$ and $\tilde\alpha$]}
The expression under the integral represents an even function of $k$ with a zero at $k=0$ and a symmetric double-peak structure. The peaks roll off at $|k|\sim \tilde\alpha^{-1/2}$
%LL , producing
which sets a UV cutoff for the integral
similar to that above for a finite-size contact model. Our numerical analysis shows that this is indeed the case, with a finite $\alpha$ translating into an effective contact size $a\approx \tilde\alpha^{1/2}$. In other words, the
%LL partial-slip
modified boundary conditions can alter the response in the proximity to the contact while rendering it unaffected at larger distances.
%LL It is straightforward to check that an alternative way of setting partial slip, $v_\parallel\propto \p_\perp v_\parallel$ (used in hydrodynamics for rough surfaces), gives qualitatively similar results.

So far we ignored the bulk momentum relaxation, setting $\gamma_p=0$ in Eq.\eqref{eq:NS}. We now proceed to show that the signatures of viscous flow identified above are robust in the presence of a background ohmic resistivity $\rho=m\gamma_p/n e^2$ so long as it is not too strong. The dimensionless parameter which governs the respective roles of resistivity vs. viscosity is
\be
\epsilon= \rho (new)^2/\eta \approx 2\gamma_{\rm ee}\gamma_p (w/v_F)^2
.
\ee
For the values $\gamma_{\rm ee}$ and $\gamma_p$ quoted above, and taking $w=1\,{\rm \mu m}$, one obtains $\epsilon\sim 10$.
Incorporating finite resistivity in the calculation is uneventful, yielding a response
\be\label{eq:V(x)_resistivity_nonzero}
V(x)=I\int \frac{dk}{\pi k}
\frac{ \rho e^{ikx} (e^{qw}-1)(e^{kw} -1) q}{q_+( e^{qw}-e^{kw})+ q_- (e^{qw}e^{kw} - 1)}
,
\ee
where $q^2=k^2+\epsilon w^{-2}$, $q_\pm=q\pm k$ (see Supplementary Information).
For $\epsilon=0$ we  recover the pure viscous result, which is negative, whereas for $\epsilon\to \infty$ Eq.\eqref{eq:V(x)_resistivity_nonzero} gives the well-known ohmic result $V(x)=\frac{\rho}{\pi}\ln\lb \coth(\pi x/2w)\rb$, which is positive. With both $\eta$ and $\rho$ nonzero, the resistance given by Eq.\eqref{eq:V(x)_resistivity_nonzero} is positive at large $x$ but remains negative close to the contact. The dimensionless threshold that determines the possibility of negative electric response depends on the actual contact size. For the values used in Fig.\,\ref{fig2} the negative response occurs  even at fairly high resistivity values corresponding to $\epsilon\lesssim 100$.

% The remarkable
The robustness of negative response can be understood by noting that viscosity is the coefficient of the highest derivative term in Eq.\eqref{eq:NS} and thus dominates at short distances $x\lesssim x_*= \sqrt{\eta/\rho(en)^2}$. The pervasive character of negative response, manifest in Fig.\ref{fig2}, will facilitate experimental detection of viscous transport. The positions of the nodes, marked by arrows in Fig.\ref{fig2}, vary with the ratio $\rho/\eta$, which provides a convenient way to directly measure the electronic viscosity.

%LL The pervasive character of negative nonlocal response can be understood by noting that viscosity, being a coefficient of the highest derivative term in Eq.\eqref{eq:NS}, dominates at short distances   $x\lesssim x_*= \sqrt{\eta/\rho(en)^2}$. Such robustness, manifest in Fig.\ref{fig2}, will facilitate experimental detection of viscous transport.

\begin{figure}
\includegraphics[scale=0.33]{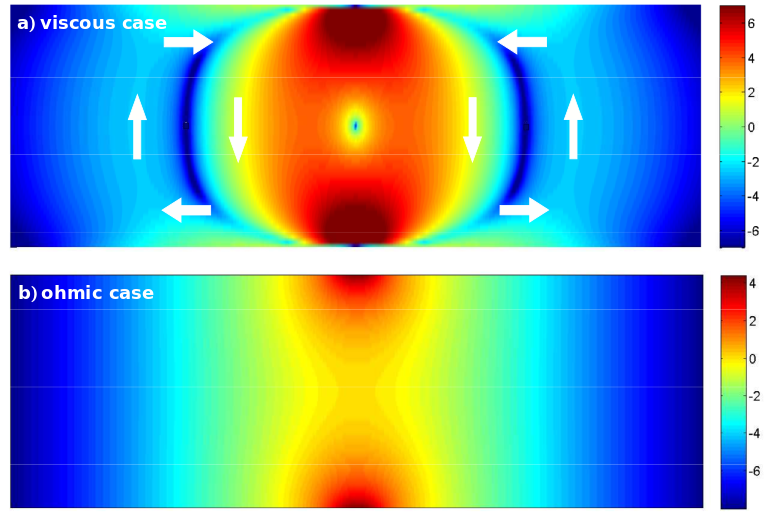}
\caption{{\bf Heating patterns for viscous and ohmic flows.} Viscous flow (panel a) results in a highly complex heating pattern with intense hot spots near contacts and cold arc-shaped patches at vortices  surrounded by warmer regions. (Also note a cold spot at the center where the flow is locally uniform and thus $W=0$, see Eq.\eqref{eq:heating_rate}). White arrows show current direction. Ohmic flow (panel b) shows an essentially featureless heat production rate decaying monotonously away from contacts.
}
\label{fig3}
%\vspace{-5mm}
\end{figure}

The hydrodynamic transport regime also features interesting thermal effects. At a leading order in the flow velocity those are dominated by convective heat transfer, described by a proportionality relation between entropy flux and flow velocity $\vec v$ (to be discussed elsewhere). At higher order in $\vec v$, besides the conventional Joule heating,
vorticity of a viscous flow
% also
manifests itself in heat production
%, which for a viscous fluid is given by
\be\label{eq:heating_rate}
W=\eta\lp\p_i v_j+\p_j v_i\rp^2=2\eta |(\p_x+i\p_y)^2\psi(x,y)|^2
\ee
The vorticity-induced heating pattern,
%for a viscous flow, presented
shown in Fig.\ref{fig3}, features hot spots near contacts and cold arc-shaped elongated patches in vorticity regions. In contrast, for an ohmic flow the pattern is essentially featureless. This makes viscous flows an interesting system to explore with the nanoscale temperature scanning techniques.\cite{T_scanning}

We finally note that there are tantalizing % possible
parallels--both conceptual and quantitative--between electronic viscous flows and current microfluidics systems.
%E.g.
Namely, a model
%an equation
essentially identical to Eq.\eqref{eq:NS} describes the low-Reynolds (microfluidic) flow between two plates separated by the distance $h$, where $\gamma_pmn=12\eta/h^2$ (it also describes viscous electron flow in a 3D slab, see Supplementary Information). A new research area at the frontier of nanoscience and fluid mechanics, microfluidics aims to manipulate and control fluids at a nanoscale with the ultimate goal of developing new lab-on-a-chip microtechnologies. Graphene, which can be easily patterned into any shapes without compromising its excellent qualities, can become a basis of  electronic microfluidics, with multiple applications in information processing and nanoscale charge and energy transport that remain to be explored.

%The authors to whom correspondence and requests for materials should be addressed: L.L. (levitov@mit.edu) and G.F. (gregory.falkovich@weizmann.ac.il).

\section{Acknowledgements}
%LL Note added:
% Note added:  After submitting our work, it came to our attention that
% negative resistance has been indeed observed in graphene (see  D. A. Bandurin et al, Science, in press,  and I. Torre, A. Tomadin, A. K. Geim, and M. Polini, Phys. Rev. B 92, 165433 (2015)).
% We thank V. F. Gantmakher, M. Yu. Reizer and D. Son for useful discussions.
%G. F. thanks A. Frishman and N. Vladimirova for help with computations;
We acknowledge support of the Center for Integrated Quantum Materials (CIQM) under NSF award 1231319
(L.L.), partial support by the U.S. Army Research Laboratory and the U.S. Army Research Office
through the Institute for Soldier Nanotechnologies, under contract number W911NF-13-D-0001 (L.L.), 
MISTI MIT-Israel Seed Fund (L.L. and G.F.), 
%This work is supported by
the  Israeli Science Foundation (grant 882) (G.F.) and the Russian Science Foundation (project 14-22-00259) (G.F.).

%\section{Author Contribution}
%
% All authors contributed to all aspects of this work.

%\vspace{-4mm}

\section{Supplementary Information}
%\appendix{\bf Supplementary Information}

\subsection{Methods}
%General remarks}
% We present here the basic tools needed
Here we outline the hydrodynamic approach used below to model electron fluids in the ohmic and viscous transport regimes. These regimes differ in the character of the constitutive current-field relation and are described by different equations for the stream function, harmonic in the ohmic case and bi-harmonic in the viscous case.
% Here we review basic mathematical facts concerning the Navier-Stokes equation in the limit of strong viscosity.
% Let us start by recalling that for

For ohmic transport the current-field relation $\vec j=\sigma\vec E$
is local, and as a result current is a potential vector field with zero vorticity. Indeed the relation %s $\vec j=\sigma\vec E$,
$\vec E=-\nabla\phi$, where $\phi$ is electrostatic potential, yields $\nabla\times\vec j =0$. Relating current density and flow velocity, $\vec j=ne\vec v$, and assuming constant particle number density $n$, we see that
the velocity field itself is potential.
Combining this with the continuity equation we write the incompressibility condition as Laplace's equation for the electric potential
\begin{equation}
\frac{ne}{\sigma}\nabla_iv_i=\nabla^2\phi=0
,\quad
\nabla^2=\partial_x^2+\partial_y^2.
\end{equation}
Incompressibility allows one to introduce the stream function via
$\vec v=\vec z\times\nabla\psi=(-\partial_y\psi,\partial_x\psi)$.
% $(v_x,v_y)=(-\partial_y\psi,\partial_x\psi)$.
The  stream function $\psi(x,y)$ in the ohmic case also satisfies Laplace's equation $\nabla^2\psi=0$.

In contrast, the current-field relation for viscous flows is essentially nonlocal. Such flows are described by the Navier-Stokes (NS) and continuity equations
\begin{equation} \eta\nabla^2 v_i= ne\nabla_i \phi,\quad\nabla_iv_i=0,
\label{S1}
\end{equation}
where $\eta$ is the dynamic viscosity. Here and below we assume the low-Reynolds limit, as appropriate for the linear response regime of interest, and consider linearized NS equation.
The velocity field of a viscous flow is generally non-potential since the vorticity $\omega=\nabla\times\vec v=\vec z \nabla^2\psi$ is  non-zero. In this case the stream function is not harmonic (unlike the Ohmic case) but rather is bi-harmonic, satisfying
% function inside the domain:
\begin{equation}
(\nabla^2)^2\psi=0
.\label{S2}
\end{equation}
We note that the bi-harmonic equation  is not conformal invariant, in contrast to the harmonic equation in the Ohmic case.
%\addQ{As discussed below, the absence of conformal invariance impacts the nonlocal response in a profound way. (?)}

The bi-harmonic equation (\ref{S2}) should be endowed with the boundary conditions which specify the behavior of the flow at system edge. One boundary condition is imposed on the normal velocity by the injected and drained current $I(r)$  flowing in and out through the leads. In terms of the stream function this reads
\be\label{eq:BC2}
\nabla_l\psi={\frac{I(r)}{en}}
.
\ee
However, since Eq.(\ref{S2}) is fourth-order, a single boundary condition would  not suffice and an extra boundary condition must be added.
%it should be endowed with extra boundary conditions.
Physically, the NS equation accounts for momentum exchange among carriers in the viscous fluid bulk, whereas the boundary condition must account for momentum exchange with the solid boundary. We start with the simplest case of the no-slip boundary condition $\vec v_l=0$ which we write as
\be\label{eq:BC1}
\nabla_n\psi=0.
\ee
Modification for the case of partial slippage is straightforward and will be discussed later. With these boundary conditions we seek the spatial dependence of the quantities $\psi$, $\vec v$, $\phi$, and then relate the net current $I$ to the potential difference at remote voltage probes $V(x)=\Delta\phi(x)$, which defines the nonlocal resistance $R_{\rm nl}(x)=V(x)/I$ (see inset in Fig.2 of the main text).

\subsection{The viscous case}
Here we describe the solution of the problem (\ref{S1}) with the no-slip boundary condition in a strip $-\infty<x<\infty$, $0<y<w$.
Using the complex variable $z=x+i y$, a general solution of the bi-harmonic equation (\ref{S2})  can be written in a compact form as $\psi(x,y)=\bar zf(z)+g(z)$. In the strip geometry the problem can be conveniently analyzed in a mixed position-momentum representation. Fourier transforming via
% In particular, for a strip of width $w$ in $y$ and infinite in $x$, we get
\begin{equation}\label{strip}
\psi(x,y)={1\over2\pi}\int dke^{ikx}\psi_k(y),
\end{equation}
the bi-harmonic equation becomes $(k^2-\p^2_y)^2\psi_k(y)=0$.

Here we consider a symmetrical lead arrangement, that is  the same current profile  $I(x)$ at $y=0,w$.
The boundary condition (\ref{eq:BC2}) for the velocity component normal to the boundary, gives
\be
v_y(x,y)_{y=0,w}=\partial_x\psi(x,y)_{y=0,w}= \frac{I(x)}{en}
,
\ee
so that $\psi_k(0)=\psi_k(w)=I(k)/en i k$. The no-slip boundary condition (\ref{eq:BC1}) reads $v_x(x,0)=v_x(x,w)=0$ giving $\partial_y\psi_k(0)=\partial_y\psi_k(w)=0$. Solving for $\psi_k(y)$ with these boundary conditions, and Fourier transforming to position space, we obtain
\bea\label{Sol1}
&&\psi(x,y)=\! \int\limits_{-\infty}^\infty\! {dk\over 2\pi}{I(k) e^{ikx}\over ne i k}\left\{{\cosh k(y - w/2)\over \cosh kw/2} \right.
\\ \nonumber
&&
\left.
+ {k\tanh kw/2\over kw + \sinh kw} [y\sinh k(w - y) + (w - y)\sinh ky]\right\}
.
\eea
%\begin{eqnarray}\label{Sol1}
%&& \psi(x,y)= {I\over ne}  \int {dk\over 2\pi}{e^{ikx}\over i k}\biggl\{{\cosh k(y - w/2)\over \cosh kw/2}+
%\\ \nonumber
%&& +\, {k\tanh kw/2\over kw + \sinh kw} [y\sinh k(w - y) + (w - y)\sinh ky]\biggr\}
%.
%\end{eqnarray}
The flow $\vec v=\vec z\times\nabla\psi$, described by this solution, can be visualized by the contours
% That formula describes the current, which flows along the lines
$\psi(x,y)={\rm const}$ which represent current streamlines.
Such streamlines are shown by white lines in Fig.\,1a of the main text. 
Notably, there are three clearly distinct groups of streamlines: one in the central region, connecting the leads along the `nominal current path', and two in the side regions, which circle around two points in the middle of the strip, $\vec r_\pm=(\pm x_0,w/2)$, where $\psi(x,y)$ peaks.  These three groups of streamlines are separated by two {\it separatrices}, defined as the outmost streamlines that border the nominal current path region  in Fig.\,1a of the main text. The two separatrices connect the right and the left edges of source and drain, respectively.

%LL The most peculiar feature of the spatial dependence given in Eq.(\ref{Sol1}) is that $\psi(x,y)$ peaks at two isolated extremum points positioned in the middle of the strip at $\vec r_\pm=(\pm x_0,w/2)$.
The streamlines circling around the extrema of $\psi(x,y)$ at the points $\vec r_\pm$ describe vortices generated by the flow.
%, as shown in Fig.\,1a of the main text.
The value $x_0$ can be obtained from the condition $\p_x\psi(x,w/2)=0$. 
%We consider first 
For the case of point-like current leads, $I(x)=I\delta(x)$, described by a constant $I(k)=I$, we have
%. Denoting $kw=2u$, we have
\begin{equation}\label{Center}
\int_{-\infty}^\infty dt\cos (2ux_0/w) \lp 1+  \frac{2u\tanh u\sinh u}{ 2u + \sinh 2u}\rp=0
,
\end{equation}
where we defined the integration variable via $kw=2u$. 
Solving this equation numerically we find $x_0\approx w$ within the numerical precision of our method.
%LL: i replaced this sentence by the one above since no separatrices show in Fig.1.
%LL Two separatrices separate the streamlines going from lead to lead from those circling around the vortex.
%The two points are the centers of the vortices seen in Figure~1 (see the main text).

The spatial dependence given by Eq.(\ref{Sol1}), evaluated for point-like leads, $I(k)=I$, is singular at $x\to 0$, $y\to 0,w$.
%, i.e at the current leads.
The singularity originates from the integrand tending to a constant value at large $k$. The large-$k$ (``ultraviolet'') regularization of the expression in Eq.(\ref{Sol1}) can be performed either by making the lead size finite or by  altering the no-slip boundary condition allowing for a small partial slip, as discussed below in Section~\ref{sec:rob}. Our numerical analysis indicates that the salient features of the flow, pictured in Fig.\,1 of the main text, are insensitive to the regularization method.

The electric potential can now be obtained from
\be
%{\partial\phi(x,y)\over\partial x}=-\frac{\eta}{ne}\nabla^2{\partial\psi(x,y)\over\partial y}
\nabla \phi(x,y)=\frac{\eta}{ne}\nabla^2\lb \vec z\times\nabla\psi(x,y)\rb
.
\label{phipsi}
\ee
The terms in $\psi$, which are purely exponential in $y$, are harmonic, that is they vanish upon applying $\nabla^2$ and do not contribute the potential $\phi$. The non-harmonic part of $\psi$ then yields
\bea\label{potentialV}
&&
\phi(x,y)={ I\eta\over \pi(en)^2} \int dke^{ikx} { k\tanh (kw/2)\over kw+\sinh kw}
\\ \nonumber
&&
\times [\sinh k(y-w)+\sinh ky]
.
\eea
The resulting spatial dependence
features a remarkable behavior which is illustrated in Fig\,1a of the main text. The potential $\phi(x,y)$, which is zero on the line $y=w/2$ by symmetry, exhibits multiple sign changes with several nodal lines separating regions where $\phi$ is positive and negative. For sufficiently small $x$, i.e. near the nominal current path line $x=0$, it varies from positive values at the source to negative values at the drain.
%grows, varying from negative values to positive values as $y$ increases.
Yet, this dependence is reversed away from the nominal current path. In particular, the potential sign {\it everywhere} at the strip boundaries $y=0,w$ outside the leads is opposite to the potential at respective leads.  This follows from the properties of the two separatrices of the flow, defined above in terms of the borders between the nominal current path region and adjacent vortex regions.
%, which connect source and drain leads. 
Consequently, every point outside the leads positioned close enough to the upper boundary is connected to the symmetrical point near the lower boundary by a streamline going opposite to the flow in the central region, which translates to a negative voltage. This can be also seen by evaluating the voltage difference for a pair of points at a distance $x$ away from the current leads:
\bea
 V(x) = && \phi(w,x)-\phi(0,x) =
%{2 I\eta\over \pi(en)^2}
\beta I \int_{-\infty}^\infty dke^{ikx} 
\\ \nonumber
&& \times {k\tanh (kw/2)\sinh kw\over kw+\sinh kw}
,\quad
\beta={2 \eta\over \pi(en)^2}
,
\label{U1}
\eea
with the parameter $\beta$ defined in the same way as in Eq.(6) of the main text. 
The voltage  $V(x)$ is a Fourier transform of a positive real-valued function which is even under $k\to-k$. To show that the resulting nonlocal resistance $R_{\rm nl}(x)=V(x)/I$ is negative, we first note that the function under the integral vanishes at $k=0$. This means that $\int_{-\infty}^\infty V(x)dx$ vanishes, and therefore $V(x)$ must be negative on a part of the domain $-\infty<x<\infty$. Next, noting that the quantity under the integral grows as $|k|$ at large $k$,
\be\label{eq:kw>>1}
\frac{k\tanh (kw/2)\sinh kw}{ kw+\sinh kw}\approx |k|
,\quad
|k|w\gg 1
,
\ee
and also taking into account the identity
\be\label{eq:identity}
\int_{-\infty}^\infty dk |k| e^{ikx}=-\frac1{(x-i0)^2}-\frac1{(x+i0)^2}
,
\ee
we conclude that $R_{\rm nl}(x)=V(x)/I$ is negative provided $x$ is nonzero and not too large.
%, has a zero at $k=0$ and grows as $|k|$ as $k\to\infty$.  We infer  that the voltage and the resulting resistance $R(x)=U(x)/I$ are  negative for any $x$.
Indeed, since at $|x|\ll w$ the integral (\ref{U1}) is determined by $|k|\gg 1/w$, the identity (\ref{eq:identity}) yields
\be R_{\rm nl}(x)\approx -\frac{2\beta}{x^2}
%-\frac{4\eta }{\pi n^2 e^2 x^2}
\label{-2}
\ee
provided $|x|\lesssim w$. Lastly, calculating the integral in (\ref{U1}) numerically we find that, as $x$ increases, $V(x)$ monotonically grows, remaining negative in the entire domain $0<|x|<\infty$ and approaching  zero without changing sign. The physical reason for the negative nonlocal voltage is a viscous vortex backflow  (see Fig.\,1a and accompanying discussion in the main text).

%\addLL{comment on the pattern of $\phi(x,y)$ sign reversals in the bulk.}

We also quote the result for the  rate of heat production  due to viscous friction:
\bea
&& Q(x,y)=\eta\sum_{i,j=x,y} (\partial_iv_j+\partial_jv_i)^2
\\ \nonumber
&&=2\eta[4\psi_{xy}^2+(\psi_{xx}-\psi_{yy})^2]=2\eta|(\p_x+i\p_y)^2\psi|^2
.\label{Heat0}
\eea
For the flow described by Eq.(\ref{Sol1}) the quantity $(\p_x+i\p_y)^2\psi(x,y)$ can be evaluated as
\bea %\label{Heat2}
&& (\p_x+i\p_y)^2\psi(x,y)={I i\over ne\pi}  \int k dke^{ikx} \left\{{e^{k(y - w/2)}\over \cosh kw/2}
\right.
\\
&&
\left.
+ {\tanh kw/2\over kw + \sinh kw}  \bigl[[1+k(y-w)]e^{ky}+(1+ky)e^{k( y - w)}\bigr]\right\}
.\nonumber
\eea
%\begin{eqnarray}
%&& (\p_x+i\p_y)^2\psi(x,y)=
%% 2I\imath  \int k dke^{ikx} \biggl\{{\cosh k(y - w/2)+\sinh k(y - w/2)\over \cosh kw/2} \label{Heat1}\\&&+ {\tanh kw/2\over kw + \sinh kw}  \bigl[ky\sinh k(w - y)+ k(w - y)\sinh ky-\cosh k(w-y)-\cosh ky\nonumber\\&&  + \sinh k(w - y) -\sinh ky-ky\cosh k(w-y)+k(w-y)\cosh ky \bigr]\biggr\} \nonumber\\&&=
%{I i\over ne\pi}  \int k dke^{ikx} \biggl\{{e^{k(y - w/2)}\over \cosh kw/2}
%\nonumber\\&&
%+ {\tanh kw/2\over kw + \sinh kw}  \bigl[[1+k(y-w)]e^{ky}+(1+ky)e^{k( y - w)}\bigr]\biggr\}
%.\label{Heat2}
%\end{eqnarray}
The resulting spatial dependence $Q(x,y)$, shown in Fig.\,3a of the main text,  is highly nonlocal.
%of the heat production is highly nonlocal as shown in Fig.\,3a of the main text}.
The heating pattern features a non-monotonic dependence with multiple cold spots and warmer regions encircling them. In particular, there are two cold spots located inside vortices and one positioned at a midpoint between the current leads.
This is in contrast to the dissipation for an ohmic flow which is maximal on the $x=0$ line and decays monotonically as $x$ increases (see Fig.3b of the main text).
This nonlocal behavior of heating can serve as another signature of viscous flows.

%, where the lower panel shows the heat production due to Ohmic resistance, $\rho (nev)^2$. The radical difference between the two maps can serve as yet another signatory, allowing one to distinguish the viscous regime with three minima (one between the electrodes and two inside vortices)
%and non-monotonic dependence on $|x|$ from the Ohmic regime with a single maximum (between the electrodes) and a monotonic decay with $|x|$. Note also the much wider hot regions around the electrodes, again manifesting nonlocality of the viscous case.

% LL STOPPED HERE

\subsection{The general ohmic-viscous case}

The approach outlined above can be generalized to describe viscous flows in the presence of ohmic resistivity. In this case, momentum transport is governed by
\be
\lp \p_t+\gamma_p\rp p_i-\eta\nabla^2v_i=-\p_i P
,
\ee
where $\gamma_p$ describes momentum relaxation due to disorder scattering. Using the incompressibility condition and introducing the stream function via $\vec v=\vec z\times \nabla \psi$ as above, we arrive at
% Let us now account for both Ohm resistance and viscosity acting together:
\begin{equation}
[\eta(\nabla^2)^2-\rho(en)^2 \nabla^2]\psi(x,y)=0
,\label{OS1}
\end{equation}
where $\rho=\gamma_p m/ne^2$ is resistivity.
%LL The same equation describes the low-Reynolds (microfluidic) flow between two plates separated by the distance $h$, where $\rho(en)^2=12\eta/h^2$.

For a strip geometry, we use the %mixed position-momentum
representation (\ref{strip}) to write  Eq.(\ref{OS1}) as
%one looks for the
% solution of (\ref{OS1}) in the form (\ref{strip}), so that (\ref{OS1}) takes the form
\begin{equation}
(\partial_y^2-k^2)(\partial_y^2-q^2)\psi_k(y)=0
,\quad
q^2=k^2+\rho(en)^2/\eta
.\label{OS11}
\end{equation}
The nonzero $\rho$ lifts the degeneracy of the eigenvalues, leading to a solution which
is a sum of four exponents:
\begin{eqnarray}\psi_k(y)=-{I\over en i k}\sum_{\pm}\bigl[ a_\pm\exp(\pm ky) +b_\pm\exp(\pm qy)\bigr]
.\label{OS2}
\end{eqnarray}
Writing the boundary conditions (\ref{eq:BC1}) and  (\ref{eq:BC2}) as
\begin{eqnarray}&& a_+e^{kw}+ a_-e^{ -{kw} }+b_+e^{qw}+b_-e^{ -qw }=1,\quad 
\\ \nonumber
&& a_++a_-+b_++b_-=1,
\\ \nonumber
&&   k(a_+- a_-)+q (b_+- b_-)=0,\quad
\\
&&
ka_+e^{kw} -ka_- e^{-{kw}}+ q b_+e^{qw} -q b_-e^{ -qw } =0
.\nonumber\end{eqnarray}
and solving for $a_\pm$ and $b_\pm$, we find
\begin{eqnarray} &&a_+= \frac{\left(e^{qw}-1\right) q}{(k-q) \left(1- e^{(k+q)w}\right)+(k+q)\left( e^{qw}-e^{kw}\right)},\nonumber\\&&
a_-= \frac{e^{kw} \left(e^{qw}-1\right) q}{(k-q) \bigl(1- e^{(k+q)w}\bigr)+(k+q)\bigl( e^{qw}-e^{kw}\bigr)},\nonumber\\&&b_+= \frac{\left(e^{kw}-1\right) k}{(q-k) \bigl(1- e^{(k+q)w}\bigr)+(k+q)\bigl( e^{kw}-e^{qw}\bigr)},\nonumber\\&&b_-= \frac{\left(e^{kw}-1\right) k e^{qw}}{(q-k) \bigl(1- e^{(k+q)w}\bigr)+(k+q)\bigl( e^{kw}-e^{qw}\bigr)}
. \nonumber
\end{eqnarray}
The potential Fourier harmonic can now be found from the relation
\begin{equation}\label{OS4}
\nabla\phi= \left(-\rho en +{\eta\over en}\nabla^2 \right)\vec v
,
\end{equation}
Plugging the general solution from Eq.(\ref{OS2})
we see that only the terms $\exp(\pm ky)$ contribute, giving
\begin{equation}\label{potential}
\phi(x,y)= {I\rho\over2\pi} \int_{-\infty}^{\infty} e^{ i kx}\Bigl[a_+(k)e^{ky}-a_-(k)e^{-ky}\Bigr] {dk\over k}
.
\end{equation}
Potential difference between the edges  of the strip $V(x)=\phi(w,x)-\phi(0,x)$ due to (\ref{potential}) is then evaluated as
\begin{eqnarray}
V(x)= {I\rho \over\pi}  \int_{-\infty}^\infty {dk\over k} \frac{e^{ikx} \left(e^{qw}-1\right)\left(e^{kw} -1\right) q}{q_+\left( e^{qw}-e^{kw}\right)+ q_-\left( e^{(k+q)w}-1\right)}.
\label{U2}
\end{eqnarray}
where $q_\pm=q\pm k$. 
This expression was used to produce Fig.2 of the main text. In doing so, we defined the dimensionless parameter
\be
\epsilon= (enw)^2\frac{ \rho}{\eta}\label{eps}
\ee
that characterizes the relative strength of the viscosity and resistivity effects, wherein the limiting values $\epsilon=\infty$ and $\epsilon=0$ correspond to the pure ohmic  and viscous regimes, respectively.

As a sanity check, we verify that in these two cases our results are in agreement with those found elsewhere. In particular,  the limit $\rho\to 0$ (i.e. $\epsilon\to 0$) can be analyzed by expanding the expression under the integral in Eq.(\ref{U2}) in a small difference $q-k=\sqrt{k^2+\epsilon/w^2}-q\sim O(\epsilon)$. Setting $\epsilon=0$ we find the resistance $R_{\rm nl}(x)=V(x)/I$ which is given by (\ref{U1}) and is {\it everywhere negative}. Similarly, in the opposite limit, $\eta\to 0$, setting $\epsilon\to\infty$ gives $qw\to\infty$, leading to
$a_\pm=(e^{\pm kw}+1)^{-1}$.
Plugging this in Eq.(\ref{potential}) leads to an expression
\begin{equation}
\phi(x,y)= {I\rho\over2\pi}  \int_{-\infty}^{\infty} e^{ i kx}
% \Bigl[\bigl(1+e^{kw}\bigr)^{-1}e^{ky}-\bigl(1+e^{-kw}\bigr)^{-1}e^{-ky}\Bigr]
%\frac{e^{ky}-e^{kw}e^{-ky}}{e^{kw}+1}
\frac{\sinh(k(y-w/2))}{\cosh(kw/2)} {dk\over k}
.
\label{potentialO}
\end{equation}
This gives the nonlocal resistance
\bea\label{U3}
&& R_{\rm nl}(x) = \frac{\phi(x,w)-\phi(x,0)}{I} 
\\ \nonumber
&& =  {\rho\over\pi}\int {e^{ikx}dk\over k} \tanh(kw/2)= {\rho\over\pi}\ln |\coth(\pi x/2w)|
\eea
which is {\it everywhere positive} and matches the result otherwise known for the ohmic regime.

%LL The dimensionless control parameter, which governs the respective roles of resistivity versus viscosity is  $\epsilon= \rho (new)^2/\eta $.

%For microfluidic flows between two plates separated by the distance $h$ the respective parameter is $\epsilon=12w^2/h^2$. We recall that for $\epsilon=0$ the resistance $R_{\rm nl}(x)=V(x)/I$ is given by (\ref{U1}), which is everywhere negative.
%In the Ohmic limit, $\epsilon\to\infty$, we have $qw\to\infty$ and  the resistance is everywhere positive: \begin{eqnarray}\lim_{\epsilon\to\infty}R =  {\rho\over\pi}\int {e^{2ikx}dk\over k} \tanh(kw)= {\rho\over\pi}\ln\left[\coth(\pi L/2w)\right].\label{U3} \end{eqnarray}

For the general case, with both $\eta$ and $\rho$ nonzero, we expect the resistance $R_{\rm nl}(x)$ to be positive at large $x$ and negative at small $x$. This behavior can be understood by putting Eq.(\ref{U2}) in the form
\bea \label{U2a}
&&
V(x)= {I\rho \over\pi}  \int_{-\infty}^\infty {dk\over k} e^{ikx} f(k)
,\quad
\\ \nonumber
&&
f(k)=\frac{e^{kw} -1}{e^{kw}+1 -\frac{k}{q}\coth(qw)(e^{kw}-1)}
.
\eea
For small enough wavenumbers $|k|\ll\sqrt{\epsilon}/w$ we have $q=\sqrt{k^2+\epsilon/w^2}\gg |k|$. In this case, since $|k|/q\ll 1$, the last term in the denominator of the expression for $f(k)$ can be dropped, giving $f(k)\approx \tanh(kw/2)$ and producing an expression for $V(x)$ which is identical to Eq.(\ref{U3}). This means that at distances larger than $\frac{w}{\sqrt{\epsilon}}$ the quantity $R_{\rm nl}(x)$ exhibits the ohmic behavior, Eq.(\ref{U3}), and is therefore positive. At the same time, for large enough wavenumbers $|k|w\gg1$ we have $qw\gg1$ and $\coth(qw)\approx 1$. In this case we can simplify the expression for $f(k)$ to read
\be
f(k)\approx \frac{\sgn k}{\epsilon} q(q+|k|)w^2
.
\ee
This expression can be used to describe the behavior of $R_{\rm nl}(x)$ at $|x|\lesssim w$. The latter is particularly simple for $\epsilon\ll 1$ (i.e. at low resistivity or high viscosity). In this case $|k|w\ll 1$ implies $q\approx |k|$, giving $f(k)\approx \sgn(k) k^2w^2/\epsilon$. Plugging this expression in Eq.(\ref{U2a}) we obtain the result identical to that in the pure viscous case, i.e. negative  $R_{\rm nl}(x)= -2\beta/x^{2}$ given in Eq.(\ref{-2}) above. A more complex behavior is found for  $\epsilon\gg 1$ (corresponding to high resistivity or low viscosity), giving $f(k)$ that behaves differently in the domains $1\ll |k|w\ll \sqrt{\epsilon}$ and $|k|w\gg \sqrt{\epsilon}$. Namely,
\be
f(k)=\begin{cases}\sgn k & 1\ll |k|w\ll \sqrt{\epsilon} \\  \frac{ \sgn k}{\epsilon} k^2w^2 & |k|w\gg \sqrt{\epsilon}
\end{cases}
%f(k)_{1\ll |k|w\ll \sqrt{\epsilon}}\approx \sgn k
%,\quad
%f(k)_{|k|w\gg \sqrt{\epsilon}}\approx \frac{ \sgn k}{\epsilon} k^2w^2
.
\ee
In the first case, after plugging in Eq.(\ref{U2a}), we find the logarithmic dependence $V(x)=(2/\pi)\rho I \ln (w/|x|)$, giving $R_{\rm nl}(x)$ of a {\it positive sign}. In the second case, we find the familiar viscous spatial dependence $R_{\rm nl}(x)= -2\beta/x^2$ of a {\it negative sign}. The two behaviors are restricted to the domains
$\frac{w}{\sqrt{\epsilon}}\lesssim |x|\lesssim w$ and $|x|\lesssim \frac{w}{\sqrt{\epsilon}}$, respectively.

The above represents a fairly dramatic behavior, which is simultaneously non-monotonic and sign-changing: as $x$ approaches the leads $R_{\rm nl}(x)$ first grows to higher and higher positive values, behaving as $R_{\rm nl}(x)=(2/\pi)\rho \ln (w/|x|)$, and then abruptly drops and reverses its sign, behaving as $R_{\rm nl}(x)= -2\beta/x^{2}$. The sign change takes place at $|x|\approx \frac{w}{\sqrt{\epsilon}}$. The complexity of this spatial dependence can be linked to the fact that viscosity represents a singular perturbation of transport equations (as a coefficient in front of the highest derivative). As a result, even when viscosity is small, it always dominates at sufficiently short distances.

Further, one can argue that the sign change of  $R_{\rm nl}(x)$ occurs at $|x|\approx \frac{w}{\sqrt{\epsilon}}$ also when $\epsilon\ll 1$ (i.e. at low resistivity or high viscosity). Indeed, the behavior at distances  $|x|\ll \frac{w}{\sqrt{\epsilon}}$ derives from wavenumbers $|k|w\gg \sqrt{\epsilon}$. In this case, $q\approx k$ for positive $k$ and $q\approx -k$ for negative $k$. Assuming without loss of generality $k>0$ and expanding Eq.(\ref{U2}) in small $q-k$  (while allowing the quantity $kw$ to be either large or small) we obtain an expression identical to Eq.(\ref{U1}) found in the pure viscous case. This expression was investigated analytically and numerically in the main text and shown to produce $R_{\rm nl}(x)$ of a negative sign.

%the resistance $R_{\rm nl}(x)$ changes sign at  $|x|\simeq \sqrt{\eta/\rho(en)^2}=\frac{w}{\sqrt{\epsilon}}$. At larger distances, $R_{\rm nl}(x)$ is positive since ohmic resistivity dominates. Indeed, the asymptotic of  (\ref{U2}) at
%$|x|\gg \frac{w}{\sqrt{\epsilon}}$ \addGF{is given by $k\ll q\approx \sqrt{\epsilon}/w$ and
%%$x\gg \sqrt{\eta/\rho(en)^2}$
%%\begin{eqnarray}\lim_{\sqrt{\epsilon} x/w\to\infty}R_{\rm nl}(x)= {\rho\over\pi}\left(1-e^{-wen\sqrt{\rho/\eta}}\right) \ln |\coth(\pi x/2w)|>0
%%.\label{U4} \end{eqnarray}
%%%\addQ{[LL: the exponent in $e^{-wen\sqrt{\rho/\eta}}$ is not dimensionless, what's the right way to fix it?]}
%%The exponential in Eq.(\ref{U4})  at $x\gg w$
%is the same as (\ref{U3}) in the purely ohmic case.}
%Viscosity dominates at smaller distances: (\ref{U2}) is close to (\ref{U1}) for $|x|\ll \frac{w}{\sqrt{\epsilon}}$
%%$|x|\ll \sqrt{\eta/\rho(en)^2}$
%so that $R_{\rm nl}(x)$ is negative.
%
%
%
%
%Viscosity is a singular perturbation (as a coefficient in front of the highest derivative).
%That means quite dramatic dependence $R_{\rm nl}(x)$ for small $\eta$: as one approaches the leads, the resistance grows to higher and higher positive values as $R_{\rm nl}(x)\propto \ln(2w/\pi x)$ and then drops fast and goes negative as $R_{\rm nl}(x)\propto -x^{-2}$  for $x\ll \sqrt{\eta/\rho(en)^2}$,  as follows from (\ref{U3}) and  (\ref{-2}).   %\addQ{[LL: what is the basis/origin of these statements?] - [GF answer: (25) and (13)]}

\subsection{The robustness of the negative nonlocal resistance}\label{sec:rob}
%effect of the boundary conditions and the contact size}

It is instructive to compare the behavior $R_{\rm nl}(x)= -2\beta/x^{2}$, which was obtained
%from the expressions (\ref{U1}), (\ref{U2})
above assuming point-like current leads and no-slip boundary conditions, to that found under more general assumptions.  One interesting generalization involves altering the boundary conditions to allow partial slippage at the boundary with the velocity proportional to the electric field:
\be\label{slip}
v_x(x,y)_{y=0}=-\partial_y\psi (x,y)_{y=0}= \alpha(x)\partial_x\phi(x,y)_{y=0}
,
\ee
and similar for $y=w$.
Here for the sake of generality we made the surface slip factor $\alpha(x)$ position dependent, which can serve as a model of structural or chemical modulation at system boundary. For a constant $\alpha$, the solution can be obtained by generalizing the approach outlined above.  This is done by seeking the stream function in the form (\ref{strip}). Potential has the same $y$-dependence as above, $\phi_k(y)=b_k[\sinh k(y-w)+\sinh ky]$, where the coefficients  $b_k$ must be found from (\ref{phipsi}) and the new boundary condition. After some algebra we find
\begin{equation}
b_k= {I\eta\over \pi(en)^2} { k\tanh (kw/2)\over kw+\sinh kw+2(\eta\alpha/ne)k^2\sinh kw}
.\label{Slip1}
\end{equation}
This expression was used in deriving Eq.(10) of the main text. The $\alpha$-dependent term in the denominator changes the large-$k$ asymptotic of $b_k$. This translates to the $x$ dependence in which the negative singularity $-1/x^{2}$ is replaced, at small enough $x$, by a much weaker positive singularity $\ln(1/x)$.
% at $x\to0$.
The value and sign of the nonlocal resistance $R_{\rm nl}(x)$ remain unaffected (and negative) at not-too-small distances $x\gg \sqrt{\eta\alpha/ne}$, indicating that  the effect of nonzero $\alpha$ is inessential and can be neglected.
We note parenthetically that it may be interesting to exploit position dependence $\alpha(x)$ by adding a small periodic modulation via $\alpha(x)=\bar\alpha+\delta\alpha\cos(kx)$. This may lead to a new behavior characterized by a modulation  $\phi(x)$ with the same periodicity, i.e. a chain of vortices.

%
%Here we probe the robustness of viscosity-induced vortices which are manifested through negative nonlocal resistance. The behavior $R_{\rm nl}(x)\propto -x^{-2}$, which was obtained from the expressions (\ref{U1}), (\ref{U2}) above, assumes point-like current leads and no-slip boundary conditions. It is instructive to compare this result to
%
%
%, feature a non-integrable

In a realistic setting, the singular behavior $R_{\rm nl}(x)\sim -1/x^2$ is also regularized by a finite contact size.
 There are two main types of contacts used in the measurements on high-mobility graphene: 
\begin{itemize}[leftmargin=*]
\item[ ] 1) ideal metal contacts;
\item[ ] 2) narrow graphene channels shaped through etching so that they connect seamlessly to the graphene bulk. 
\end{itemize}
The effect of a spread-out current 
%This can
be modeled by taking the
%\addGF{Consider now
current $I(x)$ to be distributed over
%. Let us discuss whether the different ways of regularization can alter the sign of $R_{\rm nl}$. One possible regularization of this singularity is to make the entering current distributed over
a finite cross-sectional area.
% It is straightforward to generalize
In this case Eq.(\ref{U1}) reads
\begin{eqnarray}
V(x) =
%{2 \eta\over \pi(en)^2}
\beta \int_{-\infty}^\infty\!\!\! dk\, I(k) e^{ikx} {k\tanh (kw/2)\sinh kw\over kw+\sinh kw}
.\label{UU}
\end{eqnarray}
For currents spread over a region of size $\ell$ we expect the potential at $|x|\gg\ell$ to remain unaffected, approximately following the $-1/x^2$ dependence as $|x|$ decreases. However, as $|x|$ approaches $\ell$, we expect the potential to reverse its sign and become positive at the contact. As a crude model, we illustrate this effect in the main text 
%of a non-point-like current $I(x)$ 
using a Lorentzian distribution, in Fourier representation  described by $I(k)=I\exp(-a|k|)$.

A more realistic model should account for the finite size and sharp edges of the contacts.
% \addGF{of the regions where current enters and exists the strip}.
Here we discuss the case of small-size
contacts
%\addGF{regions}
such that $\ell\ll w$,  a practically relevant case which is fairly straightforward to analyze. For $x\ll w$, using the $|k|w\gg1$ asymptotic given in Eq.(\ref{eq:kw>>1}) and the identity (\ref{eq:identity}), we can re-write Eq.(\ref{UU}) in the position space as follows:
\bea
V(x) &=&
% {2\eta\over \pi n^2e^2}
\beta \int_{-\infty}^\infty\!\!\! dk\,I(k) e^{ikx} |k|
\\ \nonumber
&=& -
% {2 \eta\over \pi n^2e^2}
\beta \int_{-\infty}^\infty dx' \lp \frac{I(x')}{(x-x'+i0)^2}+{\rm c.c.}\rp
%{k\tanh (kw/2)\sinh kw\over kw+\sinh kw}
.\label{eq:V(x)_contact}
\eea
At first, with the model 2 in mind, we ignore the equipotential condition and study potential obtained from a fixed current distribution. The resulting behavior can be exemplified by a current density constant within the region $-\ell<x<\ell$, representing contact. In this case Eq.(\ref{eq:V(x)_contact}) predicts voltage
\be\label{Vx_model}
V(x) = - \frac{2\beta}{x^2-\ell^2}I
%{2\eta I \over \pi n^2e^2(x^2-\ell^2)}
\ee
% The voltage $V(x)$
which is negative at $|x|>\ell$ and reverses sign at the edges $x=\pm \ell$; the second-order pole $x^{-2}$ found above for point contacts is now split into two first-order poles at $x=\pm \ell$.
The large-$x$ asymptotic $1/(x^2-\ell^2)=1/x^2+O(\ell^2/x^4)$ then yields the dependence far outside contacts,
\be\label{eq:ell<x<w}
V(\ell\ll |x|\ll w)=
% -{4 I\eta\over \pi (en)^2x^2}
-\frac{2\beta}{x^2} I
,
\ee
that matches the expression found for point-like leads. Deviation from this behavior occurs in the small space region $x\sim\ell$, as expected.

One can further smoothen the singularity in $V(x)$ by making the current vanish at the edge. For example, this is the case for the parabolic distribution
\be \label{eq:leads}
I(x)_{y=0,w}=\begin{cases}
3I(\ell^2-x^2)/4\ell^3,  & |x|\le\ell\\ 0, & |x|\ge\ell
%,\quad
%v_y(|x|\ge\ell)_{y=0,w}=0
\end{cases}\ .
\ee
After Fourier transforming we find $I(k)=6I[\sin(k\ell)-k\ell\cos(k\ell)]$.  Since $I(k)$ tends to a constant $I$ when $k\ell\to0$, the behavior at $x\gg\ell$ remains unaffected by the details of current distribution at the leads. In the case (\ref{eq:leads}),  integral (\ref{eq:V(x)_contact}) is conveniently evaluated by writing $\frac1{(x-x'+i0)^2}=\frac{d}{dx'}\frac1{x-x'+i0}$ and moving the derivative on $I(x')$ via by parts integration:
\bea
&& V(x) =
% {2 \eta\over \pi  (en)^2}
\beta \int_{-\infty}^\infty dx' \lp \frac{I'(x')}{x-x'+i0}+{\rm c.c.}\rp
\\ \nonumber
&& =
%-{3 I\eta\over \pi (en)^2\ell^3}
-\frac{3\beta I}{2\ell^3}\int_{-\ell}^\ell dx' \lp \frac{x'}{x-x'+i0}+{\rm c.c.}\rp
.
\eea
%The resulting integral is calculated straightforwardly
% This expression can be evaluated using
Integrating over $x'$ with the help of the indefinite integral
$\int du\frac{u}{a-u}=\int du\frac{u-a+a}{a-u}=-u-a\ln(a-u)$
%This 
gives a closed-form expression for $V(x)$ which is valid both outside and inside the leads:
\be\label{eq:V(x)_leads}
V(x) =\frac{3\beta I}{\ell^3}
%{6 I\eta\over \pi(en)^2\ell^3}
\lp 2\ell+  x\ln\frac{|x-\ell|}{|x+\ell|} \rp
%\int_{-\ell}^\ell dx' \lp \frac{x'}{x-x'+i0}+{\rm c.c.}\rp
.
\ee
The behavior at large $x$, obtained from the asymptotic $  x\ln\frac{x-\ell}{x+\ell}=-2\ell-\frac{2\ell^3}{3x^2} +O(\ell^5/x^4)$, agrees with the dependence far outside the leads found above, see Eq.(\ref{eq:ell<x<w}).
%\be
%V(\ell\ll |x|\ll w)=
%% -{4 I\eta\over \pi (en)^2x^2}
%-\frac{2\beta I}{x^2}
%,
%\ee
%that matches the expression found for point-like leads. Deviation from this behavior occurs in the small space region $x\sim\ell$, as expected.

Notably, the voltages given in Eq.(\ref{Vx_model}) and Eq.(\ref{eq:V(x)_leads}) are not constant inside the interval $-\ell<x<\ell$.  This behavior, which is physical for model 2, does not describe ideal metal contacts (model 1). In the latter case we expect the potential to be constant within the leads. To satisfy the equipotential condition one has to find the current distribution $I(x)$ and potential $V(x)$ self-consistently, in a way that ensures that the resulting $V(x)$ does not vary  inside the leads. This can be accomplished by treating 
%\addGF{This behavior might correspond to a current entering and exciting the strip via channels, but it is unphysical for current-carrying leads, which must be equipotential. To find self-consistently the current distribution, which ensures that  the voltage inside the interval $-\ell<x<\ell$ is constant and positive, one     treats}
the relation between the
 potential and current,
\be\label{eq:int_eqn}
V(x)=\beta\int_{-\infty}^\infty dx' \lp \frac{I'(x')}{x-x'+i0}+{\rm c.c.}\rp
,
\ee
as an integral equation for the function $I(x)$ localized in the  interval $-\ell<x<\ell$.
%which must be determined from the

Solution of this integral equation under the condition that $V(x)$ takes a constant value $V_0$ for all $-\ell<x<\ell$  can be found by making use of an auxiliary electrostatic problem, chosen so that the potential $V(x)$ and the derivative of the current $I'(x)$ translate into the electrostatic field and charge density, respectively.
%, can be found by making connection to electrostatics.
To that end, we consider an ideal conducting strip of width $2\ell$ in 3D placed in a uniform external electric field $\vec E_0=\lambda \hat{\vec x}$. The strip is taken to be infinite-length and zero-thickness, and positioned in the $y=0$ plane, such that
% oriented along the $z$ axis:
\be
-\ell<x<\ell,\quad y=0,\quad -\infty<z<\infty
.
\ee
The electrostatic potential spatial dependence, which includes the contributions due to the external field $\vec E_0$ and the charges it induces on the strip, can be written as
\be\label{eq:phi_strip}
\phi(x,y)=-\Re \lambda (\zeta^2-\ell^2)^{1/2}
,
\ee
where we introduced a complex variable $\zeta=x+iy$. This expression describes a function that takes a zero value within the strip $-\ell\le x\le\ell$ and has an asymptotic behavior of the form $\phi(x,y)\approx -\lambda x$ at distances large compared to $\ell$. The charge density on the strip can be found from Eq.(\ref{eq:phi_strip}) with the help of Gauss' law, giving
\be
\sigma(x)=\frac{\lambda x}{2\pi(\ell^2-x^2)^{1/2}} 
\begin{cases}1, & |x|<\ell \\ 0, & |x|>\ell
\end{cases}
.
\ee
Potential due to $\sigma(x)$ 
%this charge distribution 
accounts for the difference between the uniform field potential $\phi_0(x,y)=-\lambda x$ and the potential given in Eq.(\ref{eq:phi_strip}). For points in the $y=0$ plane this gives
\be
\int_{-\infty}^\infty \sigma(x') 2\ln\frac1{|x-x'|} dx'=\lambda x-\Re \lambda  (x^2-\ell^2)^{1/2}
,
\ee
where the branch of $(x^2-\ell^2)^{1/2}$ at negative $x$ is found by analytic continuation from positive $x$, giving $(x^2-\ell^2)^{1/2}\sgn x$.
Taking the derivative with respect to $x$ we find the electric field $x$ component $E_x(x)_{y=0}$, which gives the relation
\be\label{eq:sigma_E_relation}
\int_{-\infty}^\infty \sigma(x') \frac2{x-x'} dx'= - \lambda + \Re \frac{\lambda |x|}{  (x^2-\ell^2)^{1/2}}
.
\ee
Comparing this to the integral equation (\ref{eq:int_eqn}) we can relate current density in the leads and the  potential $V(x)$. In doing so we note that the right-hand side of Eq.(\ref{eq:sigma_E_relation}) is constant  for $-\ell<x<\ell$ since the last term in (\ref{eq:sigma_E_relation}) vanishes for such $x$. We therefore see that a solution of Eq.(\ref{eq:int_eqn}) can be obtained by identifying $\sigma(x)$ with the quantity $ - \beta I'(x)$, and $\lambda$ with the potential $V_0$ value within the lead.

Once the correspondence between the two problems is established the distribution of current in the lead can be found by solving the equation
\be
\beta I'(x)=-\sigma(x)
.
\ee
Integrating over $x$ gives the current density which is of a constant sign  within the lead and vanishes at the lead edges:
\be
I(x)=\frac{\lambda}{2\pi\beta}  (\ell^2-x^2)^{1/2}
,\quad
-\ell<x<\ell
.
\ee
Evaluating the net current as $I_0=\int_{-\ell}^\ell I(x)dx=\beta^{-1}\lambda \pi\ell^2/2$ and setting $\lambda=V_0$, we obtain the contact resistance of the lead
\be
R_{\rm c}=\frac{V_0}{I_0}=\frac{4\beta}{\ell^2}
.
\ee
The inverse-square relation $R_{\rm c}\propto \ell^{-2}$ is sharply distinct from the log dependence  $R_{\rm c}\propto \ln(w/\ell)$ found in the ohmic regime, and can therefore serve as a hallmark of the viscous regime.
The potential outside the leads can now be found simply as the right-hand side of Eq.(\ref{eq:sigma_E_relation}):
\bea\label{lead}
&& V(|x|>\ell)=V_0-\frac{V_0 |x|}{(x^2-\ell^2)^{1/2}}
\\ \nonumber
&&=
%4\beta I_0 \frac{ (x^2-\ell^2)^{1/2}-|x|  } { (x^2-\ell^2)^{1/2}}
-\frac{4\beta}{(x^2-\ell^2)^{1/2}((x^2-\ell^2)^{1/2}+|x|)  } I_0
.
\eea
The large-$x$ asymptotic matches the dependence  $V(x)=-2\beta I_0/x^2$, as expected. Interestingly, $V(x)$ takes a constant positive value $V_0$ within the source lead, while being negative outside the lead and exhibiting a square-root divergence as $x$ approaches the edges $x=\pm\ell$. The behavior on the drain side is opposite to that on the source side. The origin of the negative divergence can be understood by considering the streamlines outside of the nominal current path region,  which represent closed loops (see Fig.\,1a of the main text). 
For example, a streamline approaching one of the edges of the source lead along the strip boundary $y=w$
%, we note that the streamlines outside of the `nominal path' region bordered by separatrices are closed loops. Any streamline approaching the lead edge along the strip boundary $y=0,w$ 
must turn and leave along the respective separatrix which borders the nominal current path region. Since the separatrices make finite angles with the strip boundaries, the closer the streamline is to the boundary, the closer it comes to the lead edge and the sharper the turn it must take. Obviously, to take a sharp turn, a large enough electric field is required. This singular behavior may change upon introducing partial-slip boundary conditions, to be discussed elsewhere.

\subsection{Three-dimensional viscous electron flow}

Here we discuss the possibility to observe current vortices and negative nonlocal response in viscous charge transport in a 3D conducting slab of a small but finite thickness. In this case, as we will see, our transport equations are isomorphic to those describing the low-Reynolds (microfluidic) flow between two plates separated by a distance $h$,
% much smaller than the plate width, $h\ll w$,
where the effective resistivity is defined by $\rho(en)^2=12\eta/h^2$. Comparing this to the dimensionless control parameter introduced above, $\epsilon= \rho (new)^2/\eta $, that governs the respective roles of resistivity versus viscosity in a strip,
%, $\epsilon= \rho (new)^2/\eta $, where $w$ is the strip width,
we see that for microfluidic flows between two plates of width $w$, separated by a distance $h$, this parameter equals $\epsilon=12w^2/h^2$.

To derive the above result, we consider viscous flow of an electron fluid in an infinite slab
% of width $w$ and a finite thickness $h$
\be
-\infty<x<\infty,\quad 0<y<w,\quad 0<z<h,
\ee
in which current is injected and drained through the leads placed at the slab edges $y=0,w$.
%Denote the vertical coordinate $z$ and
%The analysis is greatly simplified for a thin slab, $h\ll w$.
We assume for simplicity that the current density at the leads has a vertical parabolic profile proportional to $z(h-z)$. In this case, the flow inside the slab is planar, $\vec v=(v_x,v_y)$, taking on a globally uniform $z$-dependence: $v_i(x,y,z)= u_i(x,y)z(h-z)/h^2$. The assumption of a parabolic $z$-dependence does not impact the generality of our analysis
%We do not loose much generality this way,
since a generic profile at the leads will transform to the parabolic profile in the vicinity of the leads, resulting in a flow in the system bulk with a parabolic $z$ dependence at the distances from the leads exceeding $h$.
The parabolic model is therefore expected to yield fully accurate results for the slab widths $w\gg h$ independent of the contact size, providing also a reasonably good approximation for $w\gtrsim h$.

To transform the 3D linearized stationary NS equation,
\begin{equation} [\eta(\partial_x^2+\partial_y^2+\partial_z^2)-\rho(en)^2] v_i= ne\nabla_i \phi,\label{3d1}\end{equation}
to a 2D NS equation, we  integrate Eq.(\ref{3d1}) over $z$ taking into account that
the potential $\phi$ is  $z$-independent because the flow is planar. Integrating the parabolic profile of the flow velocity over the cross-section then yields
% and integrate it over $h$:
\begin{equation}\label{3d3} 
[\eta(\partial_x^2+\partial_y^2)-\rho(en)^2-12\eta /h^{2}] u_i= 6ne\nabla_i \phi
.
\end{equation}
% where the potential $\phi$ is  $z$-independent because the flow is planar.
Taking the two-dimensional curl of (\ref{3d3}) we obtain transport  equations identical to those for the mixed ohmic-viscous model,  Eq.(\ref{OS1}), albeit with a renormalized resistivity:
\be
\rho(en)^2\rightarrow\rho(en)^2+12\eta /h^{2}
.
\ee
It is instructive to compare this with the dimensionless threshold that determines the possibility of negative electric response. For the contact size parameter used to produce Fig.2 of the main text, we found the threshold value $\epsilon=\rho(enw)^2/\eta\approx 120$. For a slab of thickness $h$ this translates into  $\rho(enw)^2/\eta+12w^2/h^2\leq 120$. This condition is somewhat more strict than that in 2D layers, however it can still be met even when the two slab dimensions are equal, $w\approx h$, e.g. in a wire.

\end{document}